\documentclass[prl,twocolumn,aps,showpacs]{revtex4-1}

\usepackage{epsfig}
\usepackage{graphicx}
\usepackage{amsmath}
\usepackage{amsfonts}

\begin{document}

\title{Anomalous isotope effect near a 2.5 Lifshitz transition in a multi-band multi-condensate superconductor made of a superlattice of stripes}

\author{Andrea Perali $^{1,2}$, Davide Innocenti $^{3}$, Antonio Valletta $^{4}$, Antonio Bianconi $^{2,5,6}$}

\affiliation{$^{1}$School of Pharmacy, Physics Unit, University of Camerino, 62032 Camerino, Italy\\
$^{2}$Mediterranean Institute of Fundamental Physics, Via Appia Nuova 31, 00040 Marino, Italy\\
$^{3}$CNR-SPIN and Dipartimento di Ingegneria Informatica Sistemi e Produzione, ``Tor Vergata'' University of Rome, Via del Politecnico 1, 00133 Roma, Italy\\
$^{4}$CNR-IMM, Sezione di Roma, Via del Fosso del Cavaliere 100, 00133 Roma, Italy\\
$^{5}$Rome International Center for Materials Science Superstripes (RICMASS) Via dei Sabelli 119A, 00185 Roma, Italy\\
$^{6}$Department of Physics, Sapienza University of Rome, P. le A. Moro 2, 00185 Roma, Italy}
\begin{abstract}

The doping dependent isotope effect on the critical temperature $(T_c)$ is calculated for multi-band multi-condensate superconductivity near a 2.5 Lifshitz transition. We focus on multi-band effects that arises in nano-structures and in density wave metals (like spin density wave or charge density wave) as a result of the band folding. We consider a superlattice of quantum stripes with finite hopping between stripes near a 2.5 Lifshitz transition for appearing of a new sub-band making a circular electron-like Fermi surface pocket. We describe a particular type of BEC (Bose-Einstein Condensate) to BCS (Bardeen-Cooper-Schrieffer condensate) crossover in multi-band / multi-condensate superconductivity at a metal-to-metal transition that is quite different from the standard BEC-BCS crossover at an insulator-to-metal transition. The electron wave-functions are obtained by solving the Schr\"odinger equation for a one-dimensional modulated potential barrier. The k-dependent and energy dependent superconducting gaps are calculated using the k-dependent anisotropic Bardeen-Cooper-Schrieffer (BCS) multi-gap equations solved joint with the density equation, according with the Leggett approach currently used now in ultracold fermionic gases. The results show that the isotope coefficient strongly deviates from the standard  BCS value 0.5, when the chemical potential is tuned at the 2.5 Lifshitz transition for the metal-to-metal transition. The critical temperature $T_c$ shows a minimum due to the Fano antiresonance in the superconducting gaps and the isotope coefficient diverges at the point where a BEC coexists with a BCS condensate. On the contrary $(T_c)$ reaches its maximum and the isotope coefficient vanishes at the crossover from a polaronic condensate to a BCS condensate in the new appearing sub-band.

\end{abstract}


\maketitle

\section{Introduction}


{\em Experiment.}
While for many years most of the mechanisms proposed for high temperature superconductivity have assumed a homogeneous lattice, recently new experimental results have shifted the theoretical research toward complex materials showing multi-band / multi-condensate superconductivity. Quantum oscillations have recently revealed the presence of one small electron pocket in the Fermi surface of cuprate superconductors related with the stripe phase \cite{LeBoeuf,Sebastian2,Laliberte,Khasanov,Boyer}. It has also  been proposed that the Fermi surface reconstruction in cuprates could arise from intrinsic effects in a doped Mott insulator \cite{sademelo,Ovchinnikov,imada,Norman1,Norman2,Chubukov,Birgeneau,hill}. These new results are leading the community to consider the new possibility of multi-band / multi-condensate superconductivity in charge density wave metals. The fundamental theoretical problems in this new scenario are similar to the superconductivity in ultra-narrow materials where the multi-band superconductivity is generated by quantum size effects due to the material lattice structure. Moreover the short range lattice structure in cuprates deviates from the simple average structure in the region of the phase diagram of cuprates  $T_c<T<T^*$, where deviations from a Fermi liquid metallic state are sizeable. Experimental fast and local structural methods, like Extended X-ray Absorption Fine Structure (EXAFS) \cite{exafs}, and X-ray Absorption Near Edge Structure (XANES) \cite{xanes1}, applied to cuprates \cite{Bianconi96,saini1} have shown a short-range atomic lattice reconstruction in cuprates below $T^*$ with the appearance of an incommensurate modulation of local lattice distortions. This incommensurate lattice modulation of the $CuO_2$ plane \cite{BianconiRossetti} is related with the lattice misfit strain between layers \cite{strain4,BianconiAgrestini}. The strain field in cuprates is measured by the contraction of the Cu-O bond distance from the equilibrium distance of 197 pm \cite{Garcia2} detected by XAN
ES and EXAFS. In fact, pseudo Jahn Teller polarons show a self organisation above a critical misfit strain\cite{BianconiSSC1,BianconiSSC2}. The 1D lattice modulation of the superconducting planes has been proposed to induce a reconstruction of the 2D Fermi surface due to a periodic potential of 1D potential 
barriers in the superconducting planes in cuprates \cite{PeraliSSC2,13} and in pnictides \cite{caivano}. Moreover, experimental evidences are accumulating for a 2.5 Lifshitz transition associated with a vanishing Fermi surface in the case of iron based  superconductors \cite{16,Liu,Kordyuk,Borisenko}, diborides \cite{Innocenti2010} and electron doped cuprates \cite{17}.

Moreover the physics of cuprates is controlled by an essential structural inhomogeneity \cite{Dagotto,Muller} that recently has been proposed to be made of scale invariant networks of superconducting grains \cite{Bianconig1,Bianconig2}. This scenario has been recently observed  by imaging lattice fluctuations using nano x-ray diffraction \cite{Fratini,Fratini2,pnas}, where each superconducting grain of the network is tuned near the shape resonance in superconducting gaps \cite{BianconiDiCastro}.

The small Fermi surface pockets in  YBa$_2$CuO$_{6+y}$ \cite{LeBoeuf,Laliberte,Khasanov} are characterised by a very low value of the Fermi energy, of the order of 35-40 meV. Remarkably, this very small 
Fermi energy $E_F$ is of the order of the maximum
value of the superconducting gap $\Delta$ measured at low temperature below 
$T_c$ $(\Delta/E_F \approx 1)$ by angle resolved photo-emission
spectroscopy (ARPES) and scanning tunnelling microscopy (STM). 


\begin{figure}[tpb]
\centering
\includegraphics[angle=0,scale=1.0]{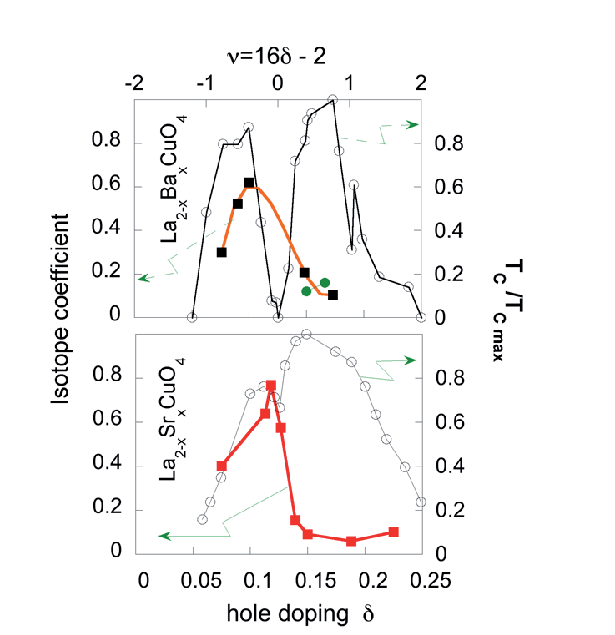}
\caption{The doping dependent isotope coefficient $\alpha$ (filled squares) and the critical temperature 
(open circles) of $La_{2-x}M_xCuO_4$ (M=Sr in the lower panel, M=Ba in the upper panel) 
as a function of hole doping $\delta$ \cite{Crawford,Farneth}. 
The isotope coefficient for $T_c$ is measured by oxygen isotope substitution, 
by replacing $^{16}O$ with $^{18}O$ and shows a pronounced 
maximum for both compounds near doping 1/8.}\label{fig1}
\end{figure}


The isotope coefficient $\alpha$ of the superconducting critical temperature is close to 0.5 in conventional BCS superconductors \cite{Maxwell} and independent on the shift of the chemical potential. This result depends on the assumption of a single Fermi surface where the Fermi level is far from the band edge and the pairing attractive mechanism is the electron-phonon Cooper pairing. The experiments clearly show that this is not the case in cuprate high-temperature superconductors, where the isotope coefficient is doping dependent and nearly zero at optimum doping \cite{Sassa,Crawford,Farneth,Zhao,Franck,Jung,Khasa,frank2,zech2,keller2,Chen}. This almost vanishing value of $\alpha$ has been considered a key evidence for an unconventional non-phononic pairing mechanism in high-temperature superconductors \cite{Zech,Kresin}. 
Here we focus our interest on the available experimental isotope effect in $La_{2-x}M_xCuO_4$ (M=Sr,Ba) \cite{Sassa,Crawford,Farneth,Zhao,Franck,Jung,Khasa,frank2,zech2,keller2,Chen} since they show a single superconducting layer with the stripe phase at 1/8.  On the contrary Y123 is a more complex bilayer cuprate with complex organization of broken oxygen chains in the basal plane \cite{keller2}.
The isotope coefficient behavior in $La_{2-x}M_xCuO_4$ (M=Sr,Ba) has a very particular doping dependence in  $La_{2-x}M_xCuO_4$ (M=Sr,Ba) \cite{Sassa,Crawford,Farneth,Zhao,Franck}, as shown in Fig. 1. It exhibits a strong anomaly clearly shown near doping 1/8 in Fig. 1, where the isotope coefficient $\alpha$ peak reaches a value close or larger than 0.5.  In these in $La_{2-x}M_xCuO_4$ systems the stripe phase is well established to appear at 1/8 doping. The large isotope effect supports the involvement of the lattice dynamics in the pairing, but in a non trivial way. The problem is very complex since there are also effects of the isotope substitution on the electronic structure \cite{Lanzara} and on the stripe phase \cite{saini1}. The isotope coefficient $\alpha$ is expected to increase near the insulator-to-metal transition, where the polaron scenario is dominant at the insulator-to-metal transition \cite{Alexandrov}, but the sharp anomaly at a particular doping indicates that the isotope coefficient shows a different anomaly related with the metallic stripe phase. It has also been shown that $\alpha$ increases The isotope effect has been discussed near a van Hove singularity in a single 2D band \cite{Markiewicz,Tsuei} and in the frame of multi-band or multi-gap superconductivity far from band edges \cite{Kristoffel,holder}, but there are missing theoretical efforts on the investigation of the anomalous isotope coefficient in a multi-band superconductor near a 2.5 Lifshitz transition at a metal-to-metal transition.


{\em Theory.}
Multi-band multi-condensate superconductivity \cite{6,7,8,holder} is usually considered for two coupled BCS condensates 
where in the normal phase the Fermi energy is far from all band edges. 
The shape resonance in the superconducting gaps \cite{6} is a type of Fano resonance between different pairing channels that occurs in multi-band metals where the chemical potential is driven at a 2.5 Lifshitz transition.
The 2.5 Lifshitz transition in the proximity of a vanishing  Fermi surface in a multi-band metal has been widely studied in the physics of Fermi surface topology in metals \cite{1,2,3,4}. The shape resonances in superconducting gaps have been shown to occur a single 2D ultra-thin metallic layer \cite{blatt} and in a metallic stripe with 1D sub-bands or mini-bands \cite{PeraliSSC1,Shanenko}. In 3D multilayer materials \cite{Bianconi94a,annette,Innocenti2010,Innocenti2011}, and in a superlattice of stripes \cite{PeraliSSC2,13} the superlattice reduces quantum fluctuations in low dimensions and the high $T_c$ coherent phase can be realised. The $T_c$ amplification is controlled by the Lifshitz energy parameter, measuring the energy difference between the chemical potential and the 2.5 Lifshitz transition. The maximum $T_c$ is reached where the Lifshitz energy parameter is of the order of the energy cut-off for the pairing interaction \cite{6,Innocenti2010,Innocenti2011}. 

In this work, we propose a model of a superlattice of quantum stripes that grabs the physical scenario emerging from the recent experiments discussed above. 
In fact, this model provides the appearing of a small electron pocket at the 2.5 Lifshitz transition as in cuprates. The fraction of the superconducting condensate originated by this small pocket has a quasi bosonic character and it is located in the crossover regime of the BCS-BEC (Bose-Einstein Condensation) crossover, a phenomenon which is deeply studied in ultracold fermions \cite{Gaebler,perali2011,Palestini}.

Therefore, our model can reproduce the formation of a quasi-bosonic condensate in the phase space 
where the small pocket appears. This type of BCS-BEC crossover
is a generic feature of multi-band / multi-condensate superconductors when the chemical 
potential is tuned close to the bottom of one of the bands and the pairing is strong enough to open gaps of the order of the 
(small) Fermi energy.


While the standard BCS-BEC crossover \cite{Leggett,Alexandrov} 
has been studied in single band metals, the present model provides a new
scenario where the BCS-BEC crossover occurs in a multi-gap superconductor. 
Here, a first BCS condensate with order
parameter $\Delta_1$ in a large Fermi surface coexists with a second Bose-like 
condensate with order parameter $\Delta_2$ in a second
small electron pocket. Pair fluctuations and their screening in the multi-gap and multi-band (or multi-patch) models
have been discussed also in the context of the physics of cuprates \cite{PeraliPRB2000,PeraliEPJB2001}.

In our model the Josephson-like coupling between the two condensates is a contact non retarded interaction due to pair exchange mechanisms, that can be attractive or repulsive. The intraband pairing is mediated by an effective attraction, having the momentum and energy structure of the electron-phonon interaction in the BCS approximation. Note that in the case of electronic mechanisms, such as exchange of spin fluctuations (paramagnons), the repulsive interaction transforms in an attractive pairing thanks to the d-wave symmetry of the superconducting order parameter, because the characteristic (large) wave-vector of the paramagnon connects state on the Fermi surface with opposite sign of the order parameter. Therefore both phononic and electronic mechanisms can be included in the model by an effective attractive interaction. 
This work is aimed to show that our simple theoretical model can explain the {\em doping dependent isotope coefficient} in cuprates near 1/8 doping in the framework of the striped phase, where the Fermi level is tuned near the 2.5 Lifshitz transition.

\section{Model and methods}

Our model is based on two well established experimental facts for the physics near 1/8 doping in cuprates : i)  a striped phase inducing Fermi surface reconstruction and 2) a 2.5 Lifshitz transition \cite{LeBoeuf,Laliberte,Khasanov}. In the underdoped regime the cuprate superconductors show at doping 1/16 a first insulator-to-metal phase transition from the doped insulating Mott phase to a correlated striped metal and at doping 1/8 a metal-to-metal Lifshitz transition for the appearance or disappearance of a small electron Fermi surface pocket coexisting with Fermi arcs. 
To investigate the response of the isotope effect at the Lifshitz transition, we consider the simplest physical model that grabs the essential physics of the cuprate metallic phase in the doping range near 1/8 doping: a 
superlattice of metallic stripes separated by a potential barrier 
\cite{Bianconi94a,Bianconi96,BianconiRossetti,BianconiSSC1,BianconiSSC2,PeraliSSC1,PeraliSSC2}
that makes a period potential in the 2D metallic layer, which provides the source for the Fermi surface reconstruction:

\begin{equation}
W(y)=\sum_{n=-\infty}^{+\infty}W_b(y-nl_p), 
\end{equation}

where $W_b(y)=-V_b$ for $\mid y\mid \leq L/2$ and $W_b(y)=0$ for
$L/2<\mid y\mid <l_p/ 2$ where $L$ is the width of the confining well
and $l_p$ is the periodicity of the superlattice in the $y$ direction.

The confining potential of Eq.(1) 
generates a band structure organised in mini-bands. 
This model allows us to simulate the electronic
structure of cuprates near the 2.5 Lifshitz transition for the appearing of a 
2D FS. The Fermi surface in cuprates is simulated by tuning the Fermi level $E_F$ 
below the bottom of the second superlattice mini-band $E_2$
where the superlattice FS (see Fig. 2) is made of a single FS of 1D character: 
the two open corrugated lines. The Fermi
surface in cuprates 
is simulated by tuning the Fermi level $E_F$ above the bottom of the second
superlattice mini-band $E_2$, where a second closed FS of 2D 
character appears beyond the first 1D mini-band (see Fig. 2).
This is determined by the quasi free electron model dispersion of the 
second mini-band $E_{2,k}^{2D}=E_2+E(k_y)+k_x^2/2m$, where
 $E(k_y)$ is the energy dispersion in the $y$ direction 
of the periodic potential $W(y)$ 
of the superlattice (here and in the following the
reduced Planck constant is set to unity). The second closed Fermi surface 
changes its 2D topology into a 1D topology with
isoenergetic open corrugated lines in the $x$ direction, above some energy 
threshold $E_{2D-1D}- E_2 =\xi$, where $\xi$ is the energy
band dispersion in the $y$ direction and $E_{2D-1D}$ is the energy where the 
topology of the FS changes from 2D to 1D.
Therefore in our model we tune 
the Fermi level in the energy range
$E_2<E_F<E_ {2D-1D}$. Multi-Gap superconductivity in the energy range 
$E_2-\xi<E_F<E_2+\xi$  requires the theoretical approach we
have recently proposed for diborides \cite{Innocenti2010} and Fe-based 
\cite{Innocenti2011} superconductors, which is capable to go beyond the standard BCS
approximations, which consist of a single band, a large Fermi surface, 
a high Fermi energy, and a constant density of states (DOS) above
and below $E_F$.


\begin{figure}[tpb]
\centering
\includegraphics[angle=0,scale=1.0]{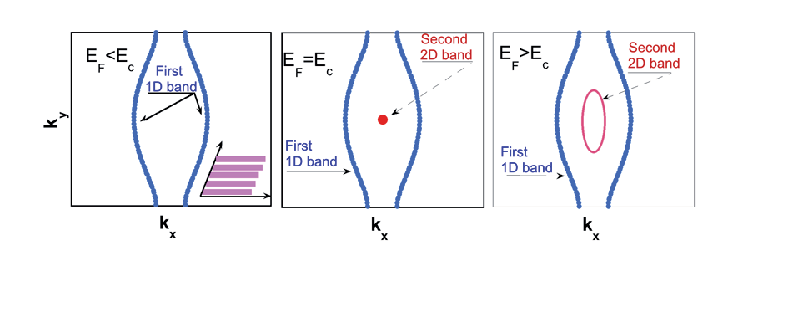}
\caption{The evolution of the Fermi surface (FS) in our model for the striped phase that 
simulates a 2.5 Lifshitz phase transition $(P_L)$ at doping $\delta=P_L=1/8$ in cuprates where a new detached electron-like Fermi surface appears. 
The inset of the left panel FS shows the lattice structure of a superlattice of stripes with finite hopping probability between stripes. 
$E_c$ is the energy of the bottom of the second mini-band. 
The chemical potential is tuned from below (left panel $\delta<P_L$) to 
above (central panel $\delta>P_L$) where a new small 2D closed electron Fermi surface 
appears (red circle). The right panel shows the FS beyond the electronic topological 
transition (ETT) of the second band from a 2D FS to a corrugated 1D FS at higher
doping called 2D-1D ETT.}\label{fig2}
\end{figure}


We consider a 1D periodic potential barrier of width B and wells of width L in 
the $y$ direction with periodicity $l_p=L+B=1.9nm$ 
to simulate the superstructure of Y-based and La-based cuprates 
and constant along  the $x$ direction. 
The potential barrier is fixed at $V_b=1400meV$. 
We note that small variations of $l_p$ and of other parameters 
of the potential do not influence the main results of this work. 
The strength of the potential barrier $V_b$ is important to determine
the 1D to 2D dimensional crossover observed in the phase diagram of
cuprates: lower values of $V_b$ will give a quasi-2D electronic system
with no interesting stripe features, while larger values of $V_b$
will lead to a strong 1D stripe physics, without the possibility of
formation of electron pockets in the Brillouin zone, and without
the screening of order parameter fluctuations active in a dimensional
crossover (pronounced 1D anisotropy suppresses in general superconducting long-range order). 
The present choice of the periodic potential gives a band 
dispersion $\xi=50meV$. The band dispersion $\xi$ is
two times the electronic hopping integral $t_y$ 
in the direction $y$, that is much 
smaller that the hopping integrals $t_x$ in the $x$
direction. Solving the Schr\"odinger equation for the 1D periodic potential 
barrier of Eq.(1), we obtain the wave-functions of the
electrons with a free electron band dispersion along the stripe direction and 
tight-binding mini-bands in the transverse
direction. The eigenvalues are labelled by three quantum number 
$E=\epsilon_{n,k_x,k_y}$ 
where $n$ is the mini-band index, $k_x$ and $k_y$ are the components of the
electron wave-vectors in the superlattice. The DOS as function of the energy
shows a jump at $E_{edge}=E_2$ and a sharp peak at $E_{2D-1D}$.
The superconducting phase occurs because of the presence of an attractive 
intraband electron-electron effective interactions
(1,1) and (2,2) in the first and second band respectively and the interband 
exchange-like interactions (1,2) and (2,1). The cutoff energies of
the interactions are symmetrically fixed around the Fermi surface and the 
value of the effective coupling has been fixed at $\lambda=1/3$.
In the BCS approximation, i.e., a separable interaction in wave-vector space, 
the gap parameter has the same energy cut off
as the interaction. Therefore it has a value $\Delta_{n,k_y}$ around the Fermi
surface in a range of energies equals to the energy cutoff,
depending from the mini-band index $n$ and the superlattice wave-vector $k_y$. 
The self consistent equation for the ground state
($T=0$) energy gap $\Delta_{n,k_y}$ is:

\begin{equation}\begin{split}
\Delta_{n,k_{y}}=-\frac{1}{2N}\sum_{n',k'_y,k'_x}
\frac{V_{\mathbf{k},\mathbf{k}'}^{n,n'}\Delta_{n',k'_y}}{
\sqrt{(E_{n',k'_y}+\epsilon_{k'_x}-\mu)^2+\Delta_{n',k'_y}^2}},
\end{split}\end{equation}

where N is the total number of wave-vectors in the discrete summation, 
$\mu$ is the chemical potential, $V^{n,n'}_{\mathbf{k},\mathbf{k}'}$ is the effective 
pairing interaction 

\begin{equation}\begin{split}
&V_{\mathbf{k},\mathbf{k}'}^{n,n'}=
\widetilde{V}_{\mathbf{k},\mathbf{k}'}^{n,n'}\\&\times\theta(\omega_0-|E_{n,k_y}+\epsilon_{k_x}-\mu|)\theta(\omega_0-|E_{n',k'_y}+\epsilon_{k'_x}-\mu|)
\end{split}\end{equation}

calculated taking into account the interference effects between the 
wave functions of the pairing electrons in the different
mini-bands, where $n$ and $n'$ are the mini-band indexes, $k_y(k_y')$ is the 
superlattice wave-vector and $k_x(k_x')$ is the component of the
wave-vector in the stripe direction of the initial (final) state in the 
pairing process, and

\begin{equation}\begin{split}
&\widetilde{V}_{\mathbf{k},\mathbf{k}'}^{n,n'}=-\frac{\lambda_{n,n'}}{N_0}S\\&\times\int_{S}
\psi_{n',-k'_y}(y)\psi_{n,-k_y}(y)
\psi_{n,k_y}(y)\psi_{n',k'_y}(y)dxdy,
\end{split}\end{equation}

where $N_0$ is the DOS at $E_F$ for a 
free-electron 2D system, $\lambda_{n,n'}$ is 
the dimensionless coupling parameter, $S=L_xL_y$ is the
surface of the plane and $\psi_{n,k_y}(y)$ are the eigenfunctions in the 
superlattice of quantum stripes. The gap equations have been 
solved iteratively. We obtain anisotropic gaps strongly dependent on 
the mini-band index and weakly dependent on the
superlattice wave-vector $k_y$. According with Leggett \cite{Leggett}, 
the ground-state BCS wave function corresponds to an ensemble
of overlapping Cooper pairs at weak coupling (BCS regime) and evolves to 
molecular (non-overlapping) pairs with bosonic
character. The point is that the BCS equation for the gap has to be 
coupled to the equation that fixes the fermion density:
with increasing coupling (or decreasing density), the chemical potential $\mu$ 
results strongly normalised with respect to the
Fermi energy $E_F$ of the non interacting system, and approaches minus half 
of the molecular binding energy of the corresponding two-body problem in the vacuum. 
Therefore in order to correctly describe the case of the chemical potential
near a band edge, where all electrons in the new appearing band condense 
forming a bosonic-like gas in the second mini-band, 
the chemical potential in the superconducting phase is normalised 
by the gaps opening at any chosen value of the
charge density $\rho$:

\begin{equation}\begin{split}
\rho &=\frac{1}{L_xL_y}\sum_{n}^{N_b}\sum_{k_x,k_y}\left[1-
\frac{E_{n,k_y}+\epsilon_{k_x}-\mu} {\sqrt{(E_{n,k_y}+\epsilon_{k_x}-\mu)^2+\Delta_{n,k_y}^2}}\right]\\&=\frac{\delta k_y}{\pi}\sum_{n=1}^{N_b}\sum_{k_y=0}^{\pi /l_p}
\int_{0}^{\epsilon_{min}}d\epsilon \frac{2N(\epsilon)}{L_x}+\int_{\epsilon_{min}}^{\epsilon_{max}}d\epsilon \frac{N(\epsilon)}{L_x}\\&\times\left(1-
\frac{E_{n,k_y}+\epsilon_{k_x}-\mu} {\sqrt{(E_{n,k_y}+\epsilon_{k_x}-\mu)^2+\Delta_{n,k_y}^2}}\right),
\end{split}\end{equation}
where
\begin{equation*}\begin{split}
&\epsilon_{min}=max\left[0,\mu - \omega_0 - E_{n,k_y} \right],\\
&\epsilon_{max}=max\left[0,\mu + \omega_0 - E_{n,k_y} \right],\\
&N(\epsilon)=\frac{L_x}{2\pi\sqrt{\frac{\epsilon}{2m}}},
\end{split}\end{equation*}
and $N_b$ is the number of the occupied mini-bands, $L_x$ and $L_y$ are
 the size of 
the considered surface and the increment in
$k_y$ is taken as $\delta k_y = 2\pi/L_y$. We compute the critical temperature
 $T_c$ 
of the superconducting transition solving the linearised BCS equations

\begin{equation}
\Delta_{n,k_y}=
-\frac{1}{2N}\sum_{n',\mathbf{k'}}V_{\mathbf{k},\mathbf{k}'}^{n,n'}
\frac{\tanh(\frac{E_{n,k_y}+\epsilon_{k_x}-\mu}{2T_c})}{E_{n,k_y}+\epsilon_{k_x}-\mu}\Delta_{n',k'_y},
\end{equation}

where the energy dispersion is measured with respect to the chemical 
potential. The iterations are stopped when a
convergence factor of $10^{-6}$ has been reached, starting with a trial 
temperature $T_1$ and finding the $T_c$ by the Newton tangent method to solve 
the implicit integral equation for $T_c$. The $T_c$ is evaluated as a function 
of the chemical potential in the
proximity of the edge of the 2-nd mini-band. The tuning of the chemical 
potential is measured by the Lifshitz parameter
$z=(\mu-E_{2})/\omega_0$ where $E_2$ is the bottom of the second band 
and $\omega_0$ is the energy cut-off for the pairing interaction.

\begin{figure}[tpb]
\centering
\includegraphics[angle=0,scale=1.0]{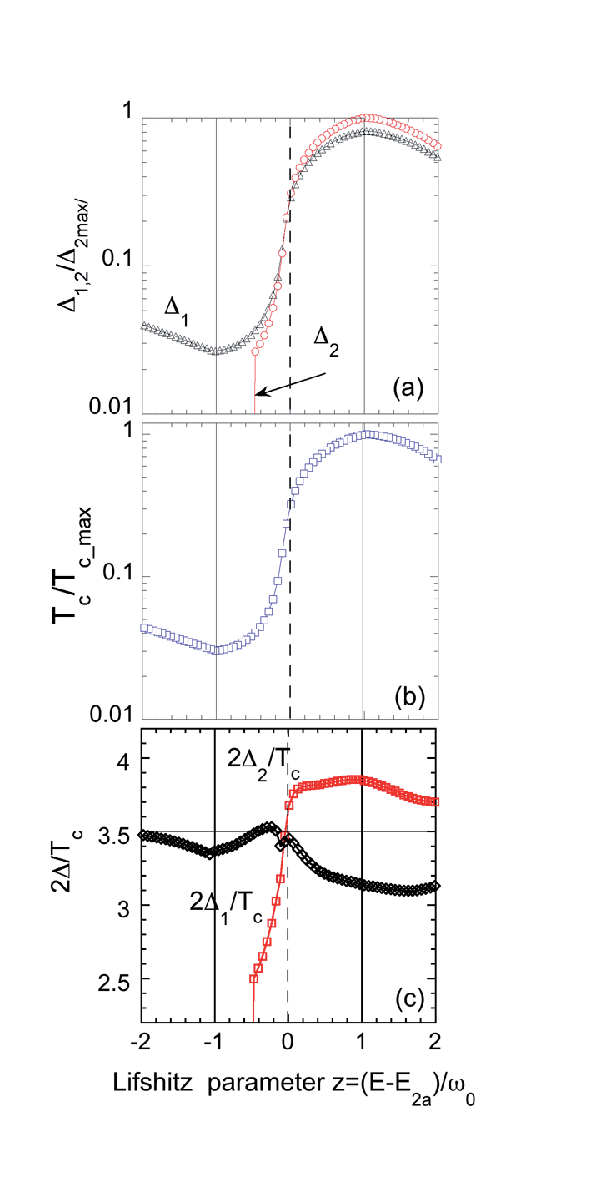}
\caption{a) The superconducting gaps in the first and second mini-bands normalised to the maximum value of the gap in the second mini-band; b) the critical temperature $T_c$ normalised to its maximum value; c) and the gaps to $T_c$ ratios for the two gaps as a function of the Lifshitz parameter $z=(\mu-E_{2})/\omega_0$ compared with the standard BCS value of 3.5.}\label{fig3}
\end{figure}

\section{The isotope coefficient}

The experimental 2.5 Lifshitz phase transition observed in the experiment reported 
by ref. \cite{LeBoeuf,Sebastian2,Laliberte} at doping $\delta=P_L=1/8$ (reduced doping $\nu=16\delta-2=0$ ) is simulated by tuning the Fermi level at $E_F=E_2$,
$i.e.$, at $z=0$, corresponding with the reduced doping range $-1<\nu<+1$. The gaps, the $T_c$ 
and the gap to $T_c$ ratios have been plotted as a function 
of the Lifshitz energy parameter in the range $-1<z<+1$, as shown in Figure 3. We obtain the minimum of $T_c$ 
where $\mu$ is tuned at $z=0$, $i.e.$, at the 2.5 Lifshitz transition
for the appearance of a new detached electron Fermi surface that 
well reproduce the minimum at 1/8 in Fig. 1. The
maximum of $T_c$ occurs where the topology of the FS of the 2-nd sub-band shows 
the 2D->1D electronic topological transition. This result is assigned to the shape resonance in the two 
superconducting gaps controlled by the
exchange-like pair-transfer (or Josephson-like) pairing that shows a minimum (maximum) of $T_c$ at the antiresonance at $z=-1$ (resonance at $z=1$) due to the negative (positive) quantum interference effects typical of the shape resonances in the superconducting gaps.

Our results well reproduce the increase of the critical temperature $T_c$ going 
from 0.125 to 0.18 doping in cuprates.
The isotope effect $\alpha=\partial ln T_c / \partial ln M$ is calculated with 
the assumption of an energy cutoff of the interaction 
dependent from the isotopic mass as $\omega\propto M^{-1/2}$. 
In Fig. 4 we report the calculated isotopic 
coefficient $\alpha$ as a function of the chemical potential that strongly
deviates from the standard BCS value $\alpha=0.5$.
 We find the maximum of $T_c$ and the minimum of $\alpha$ at the shape resonance $\mu=E_{2D-1D}$, because the $T_c$ is weakly dependent by small variation of the energy cutoff.
On the contrary by tuning the chemical potential at the 2.5 Lifshitz transition for the appearance of the second circular
Fermi surface, $i.e.$, at the band edge of the second mini-band $\mu=E_2$, we find a large value of $\alpha$ $\gg 0.5$ and a drop of $T_c$ in agreement with the experimental data. The isotope coefficient has been 
calculated considering the cases where only one of
the intraband pairing energies $\omega_{11}\propto M^{-1/2}$ or
 $\omega_{22}\propto M^{-1/2}$ and the interband pairing energy
 is isotope dependent or independent. The best agreement with 
the experimental data reported in Fig. 4 is obtained for the case where
both intraband pairing energy $\omega_{11}\propto M^{1/2}$ and $\omega_{22}\propto M^{1/2}$
are isotope dependent and the
electronic interband energy is isotope independent. Note that at the 
Lifshitz phase transition at $(z=0)$ both $\alpha$ and the gap to
$T_c$ ratios in both bands get exactly the conventional BCS values. 
This is the crossing point between the Bose-like regime
where $\Delta_2$ is smaller than $\Delta_1$ but larger than $\mu-E_2$
($i.e.,$ the Fermi surface in band 2 is destroyed by the gap opening)
and the BCS-like regime where $\Delta_2$ is larger than $\Delta_1$ but smaller than $\mu-E_2$ ($i.e.$, the Fermi surface in band 2 is only partially smeared by the gap opening).

\begin{figure}[tpb]
\centering
\includegraphics[angle=0,scale=1.0]{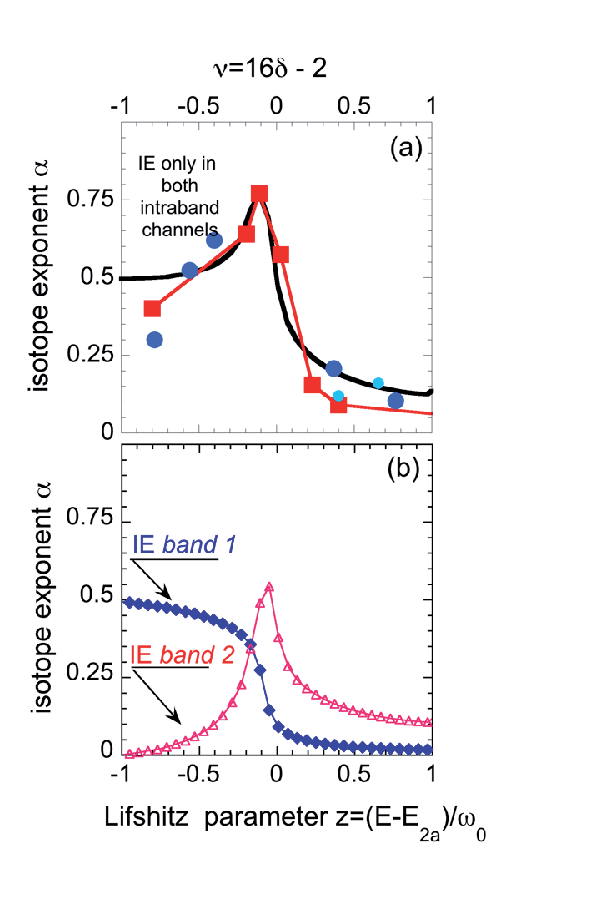}
\caption{panel a: The experimental isotope coefficient $\alpha$ as a function of the reduced 
doping $\nu=16\delta-2$ for $La_{2-x}M_xCuO_4$ (M=Sr dots and M=Ba squares) compared with the theoretical isotope coefficient $\alpha$ as a function of the Lifshitz parameter for the case of an isotope effect (IE) driven by intraband pairing in both bands $\omega_{11}\propto M^{1/2}$ and $\omega_{22}\propto M^{1/2}$ while the Josephson-like exchange-like pair transfer mechanism is isotope independent; panel b) theoretical isotope coefficient $\alpha$ as a function of the Lifshitz parameter for the case of an isotope effect (IE) only the intraband pairing energy in band 1 $\omega_{11}\propto M^{1/2}$ (filled dots) and for the case of an isotope effect (IE) only the intraband pairing energy in band 2 $\omega_{22}\propto M^{1/2}$ (empty triangles).}\label{fig4}
\end{figure}

\section{Discussions and Conclusions}

Starting from the stripes scenario near 1/8 for the electronic structure of 
cuprate superconductors and the identification of
the 2.5 Lifshitz transition in the phase diagram of cuprates, 
we study the behaviour of the critical temperature $T_c$ and the isotope 
coefficient of the superconducting phase transition 
as a function of the chemical potential near the 2.5 Lifshitz metal-to-metal transition,
which is located at the band edge of a sub-band for a superlattice of quantum stripes. 

We obtain an interesting asymmetric feature of the 
chemical potential dependence of $T_c$ and $\alpha$. This peculiar shapes of
$T_c$ and $\alpha$ given by our numerical calculations
are well in agreement with the phenomenology of cuprate superconductors 
near doping 1/8. The energy cutoff $\omega_0$ of the effective
pairing interaction considered in our model determines the width of the
shape resonance in the superconducting gaps and the small value of the isotope coefficient 
in the flat region of its doping dependence, well below the standard
BCS value of $\alpha$=0.5, where $T_c$ has a maximum in agreement with experiments.
We find the maximum $T_c$ when the chemical potential is tuned near the 2D-1D electronic topological transition 
of the 2-nd sub-band (where the Lifshitz
parameter assumes the value $z=1$). The maximum
value of the isotope coefficient is reached at the 2.5 Lifshitz transition, 
in the range of the Fano antiresonance where $T_c$ is strongly suppressed. 
Interestingly, the ratios between the gaps and $T_c$ in different sub-bands cross the conventional BCS
ratio values (=3.5) at the 2.5 Lifshitz transition, while sizable deviations from the conventional value are obtained in our calculations above this transition, in the range $1<z<2$ accessible to experiments.

Therefore, we have provided a simple model that reproduces the enhancing of 
the isotope coefficient well above the conventional BCS value and
explains the rapid doping dependence of the isotope coefficient $\alpha$
observed in the $La_{2-x}M_xCuO_4$ (La214) and $YBa_{2}Cu_{3}O_{6+y}$ (Y123) systems near doping 1/8. The present results have an impact on the physics of superconductivity in nano-sized superconductors \cite{Zech,Kresin,Shanenko}, where the shape resonance in superconducting gaps is gaining momentum as a key ingredient for the road map for new high-$T_c$ superconductors.
Moreover, shape resonances and quantum size effects have been also considered
in the context of superconductivity in nanofilms \cite{ShanenkoPerali1}
and superfluidity in cigar-shaped ultracold Fermi gases \cite{ShanenkoPerali2} as a possible new driving mechanism to tune 
an atypical BCS-BEC crossover in multi-band fermionic systems, with the appearance of a coherent mixture of 
BCS-like and BEC-like condensate. Finally, we have shown that the anomalous maximum of the isotope effect and its peculiar doping dependence in the doping range 1/8, shown in Fig. 1 in La214 families, could have its origin in a 2.5 Lifshitz transition for a metal-to-metal transition in a multi-band multi-condensate superconductor, for the appearing or disappearing of small electron-like Fermi surface pocket. Therefore, the experimental investigation of the complex doping dependent isotope coefficient $\alpha$ in the 1/8 doping range in other cuprate and pnictides families of superconductors could provide a signature for the role of the 2.5 Lifshitz transitions in the superconducting phase, in agreement with a large Nernst effect measured in the normal phase \cite{2,3}.



\end{document}